\documentclass[sigconf]{acmart}
\usepackage{subcaption}
\usepackage{pifont}
\usepackage{balance}
\setcopyright{none}
\renewcommand\footnotetextcopyrightpermission[1]{}
\AtBeginDocument{%
  }

\copyrightyear{2025}
\acmYear{2025}
\setcopyright{acmlicensed}\acmConference[*]{*}{*}

\acmBooktitle{-}

\begin{document}

\title{Exploring Palette based Color Guidance in Diffusion Models}

\author{Qianru Qiu}
\email{qiu\_qianru@cyberagent.co.jp}
\affiliation{%
  \institution{CyberAgent Inc.}
  \city{Shibuya}
  \state{Tokyo}
  \country{Japan}
}

\author{Jiafeng Mao}
\email{jiafeng\_mao@cyberagent.co.jp}
\affiliation{%
  \institution{CyberAgent Inc.}
  \city{Shibuya}
  \state{Tokyo}
  \country{Japan}
}

\author{Xueting Wang}
\email{wang\_xueting@cyberagent.co.jp}
\affiliation{%
  \institution{CyberAgent Inc.}
  \city{Shibuya}
  \state{Tokyo}
  \country{Japan}
}


\begin{abstract}
  With the advent of diffusion models, Text-to-Image (T2I) generation has seen substantial advancements. Current T2I models allow users to specify object colors using linguistic color names, and some methods aim to personalize color-object association through prompt learning. However, existing models struggle to provide comprehensive control over the color schemes of an entire image, especially for background elements and less prominent objects not explicitly mentioned in prompts. This paper proposes a novel approach to enhance color scheme control by integrating color palettes as a separate guidance mechanism alongside prompt instructions. We investigate the effectiveness of palette guidance by exploring various palette representation methods within a diffusion-based image colorization framework. To facilitate this exploration, we construct specialized palette-text-image datasets and conduct extensive quantitative and qualitative analyses. Our results demonstrate that incorporating palette guidance significantly improves the model’s ability to generate images with desired color schemes, enabling a more controlled and refined colorization process.
\end{abstract}

\begin{teaserfigure}
  \includegraphics[width=\textwidth]{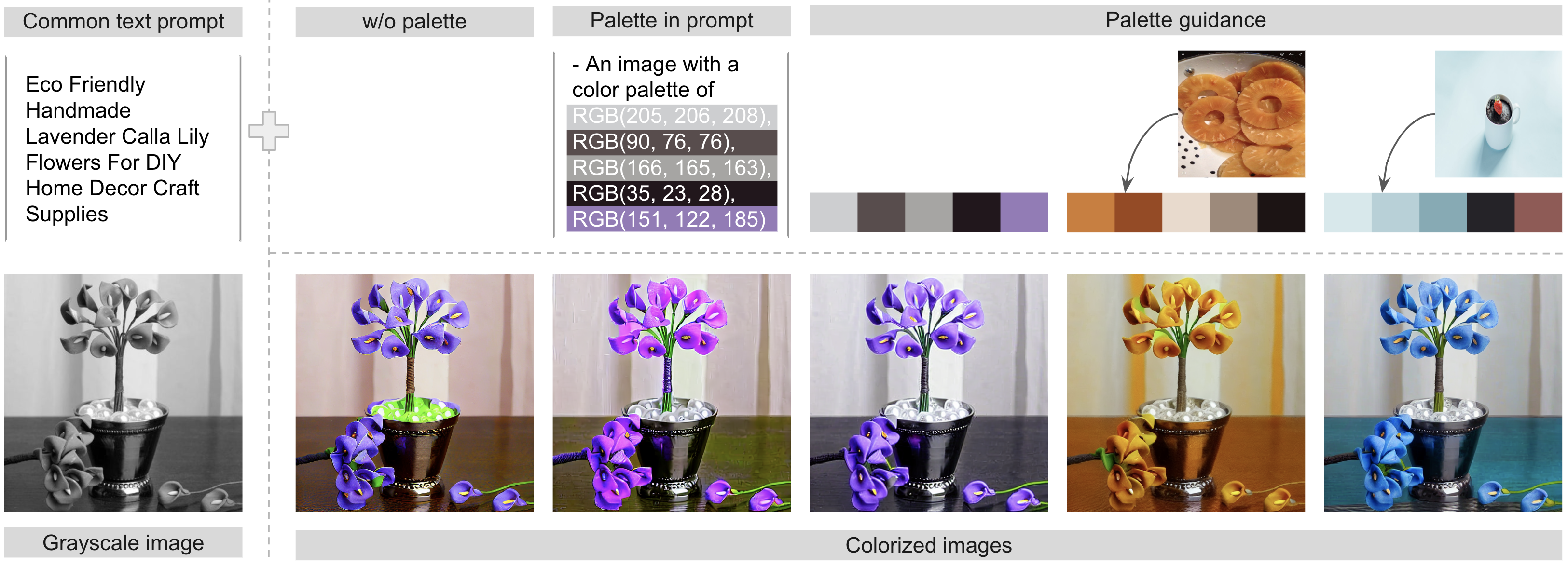}
  \caption{Colorized results produced by diffusion models under palette guidance conditions. The grayscale image and text prompt in the left column serve as the common conditions for all colorized images. The first colorized image is generated using only the common text prompt, resulting in the flowers being colorized with a cue of lavender color, while the colors of other parts lose control. In the second image, additional RGB values are incorporated directly into the text prompt, but the resulting colors of flowers and background deviate from the specified colors. The next three images are generated using our proposed palette guidance, which align more precisely with the specified colors. The first of these uses a user-selected palette and the other two employ palettes corresponding to the reference images.}
  \label{fig:overview}
\end{teaserfigure}


\maketitle

\section{Introduction}
\label{sec:intro}

Color control plays a crucial role in image generation, as it significantly enhances the diversity and controllability of generated images. Effective color management allows users to achieve specific aesthetic goals and ensures that the output aligns with their intended visual expression. 
In current diffusion-based T2I models, users typically specify colors using linguistic terms like “red” or “blue”. These general color terms often lead to imprecise outputs that may not match the user's exact preferences, especially in professional color design contexts. Even if users attempt more precise color descriptions or directly embed specific RGB values into text prompts, existing methods primarily focus on specifying color-object pairs within prompts~\cite{rombach2022high, podell2023sdxl, weng2024cad, atif2024colorpeel}. 
However, these models struggle to control the overall color scheme of an entire image, particularly for elements not explicitly referenced in the text prompt, such as backgrounds or less prominent objects.
This limitation often results in inconsistencies in the color harmony of generated images, as demonstrated by the light green color in the first colorized image of Figure~\ref{fig:overview}.

To address these limitations and enhance color control, we propose the use of color palettes as a distinct guidance mechanism, separate from text prompts. A color palette is a refined set of colors that captures the primary colors of an image. Palettes are effective tools for achieving consistency and harmony in color management, especially in professional design work. It has been widely used in various applications, including 2D image recoloring~\cite{tan2018efficient, chang2015palette, cho2017palettenet}, video recoloring~\cite{du2021video}, and color editing of 3D scenes~\cite{gong2023recolornerf, kuang2023palettenerf}. Despite the importance of palettes in precise color management, how to effectively integrate them as a guidance modality in diffusion models remains unexplored. 

Our work aims to bridge this gap by introducing palette guidance into generative diffusion models for improved color control. The main challenge lies in effectively integrating palette information into the image generation pipeline. A straightforward approach, directly inserting a palette's raw color values (e.g., RGB) into text prompt, proves ineffective, as the CLIP~\cite{radford2021learning} text encoder commonly used in T2I models is not well-suited for interpreting explicit color values.
This naive approach can lead to unintended color outputs, as demonstrated in the second colorized image of Figure \ref{fig:overview}.

To overcome these challenges, we explore multiple strategies for effective palette representation and develop a separate palette representation model to produce learned palette embeddings. These palette embeddings are then fused with text embeddings to guide the image generation process. We investigate different modalities for training the palette representation models, and adopt the image colorization task as a controlled setting to verify the effectiveness of our palette guidance on color control. This application allows users to either customize their own palettes or use palettes derived from reference images to control the overall color themes of the generated images, as illustrated on the right side of Figure \ref{fig:overview}.

In summary, our main contributions include:
\begin{itemize}
\setlength{\itemsep}{0mm}
    \item We propose a novel approach to enhance control over the color schemes of generated images by integrating color palette guidance into diffusion-based generative models.
    \item We investigate the impact of using different modalities for palette representation, providing insights into their effects on color control in generated images.
    \item We conduct extensive evaluations, including quantitative experiments and a user study with professional designers, which validated the effectiveness of our method and confirmed the practical demand for palette-based guidance in generation tasks.
\end{itemize}
\section{Related work}
\label{sec:relatedwork}

\subsection{Color palette representation}

Early efforts in color palette representation~\cite{o2011color, kita2016aesthetic} employed regression models with handcrafted feature extraction to learn relationships among colors in a palette. These methods were limited by their dependence on manually designed features, which often struggled to capture the complex and nuanced relationships between  palette colors. Recently, researchers have explored text embedding techniques for palette representation, essentially treating individual colors as words and entire palettes as sentences. For instance, Kim et al.~\cite{kim2022colorbo} trained a color embedding model for Mandala Coloring in a manner similar to fastText~\cite{bojanowski2017enriching} which provides continuous vector representations of color words, preserving semantic similarity by encoding their meanings in a vector space. Qiu et al.~\cite{qiu2023color, qiu2023multimodal} further proposed a Transformer-based masked color model for multi-palette representation. This model leverages an attention mechanism to capture relationships between colors within a palette and their connections to textural contexts. Our work extends these methods to explore how palette representations can be conditioned on multiple modalities, such as text and image inputs.

\subsection{Diffusion-based image colorization}

We employ the image colorization task as a controlled use case to evaluate the effectiveness of palette guidance in enhancing color controllability. Image colorization is the task of adding colors to grayscale images. To enable customized colorization, various conditional methods have utilized different user-provided inputs, such as strokes~\cite{xiao2019interactive}, reference images~\cite{xu2020stylization, yin2021yes}, and text descriptions~\cite{chang2022coder, weng2022code, weng2024cad}.
Recent advances in diffusion models have led to significant improvements in conditional colorization, and there is increasing interest in text-driven colorization, where users specify detailed color descriptions for individual objects in an image. An early work, Diffusing Colors~\cite{zabari2023diffusing} introduced a diffusion-based framework that conditions generation solely on text. More recently, L-CAD~\cite{weng2024cad} leveraged latent diffusion models for image colorization using flexible textual descriptions. However, text-based methods often focus on specifying object-color pairs in the prompt, leaving other elements with limited or imprecise color control. To address the limitations of relying solely on text input for colorization, we propose incorporating color palettes as an additional control modality. Notably, ControlNet~\cite{zhang2023adding} has emerged as a powerful architecture for conditioning diffusion models, capable of flexibly accepting various types of conditions, such as edges or depth maps. In the context of image colorization, ControlNet can be trained on paired grayscale-color images, enabling it to learn how to propagate the spatial structure from the grayscale input. In our evaluation, we compare our method with L-CAD and ControlNet using a grayscale image as a common conditioning input. Since L-CAD demonstrated that diffusion-based models outperform GANs- and Transformers-based models for text-driven image colorization, we do not include comparisons to earlier non-diffusion methods.

\begin{figure*}
  \centering
  \begin{subfigure}{0.61\linewidth}
    \includegraphics[width=\linewidth]{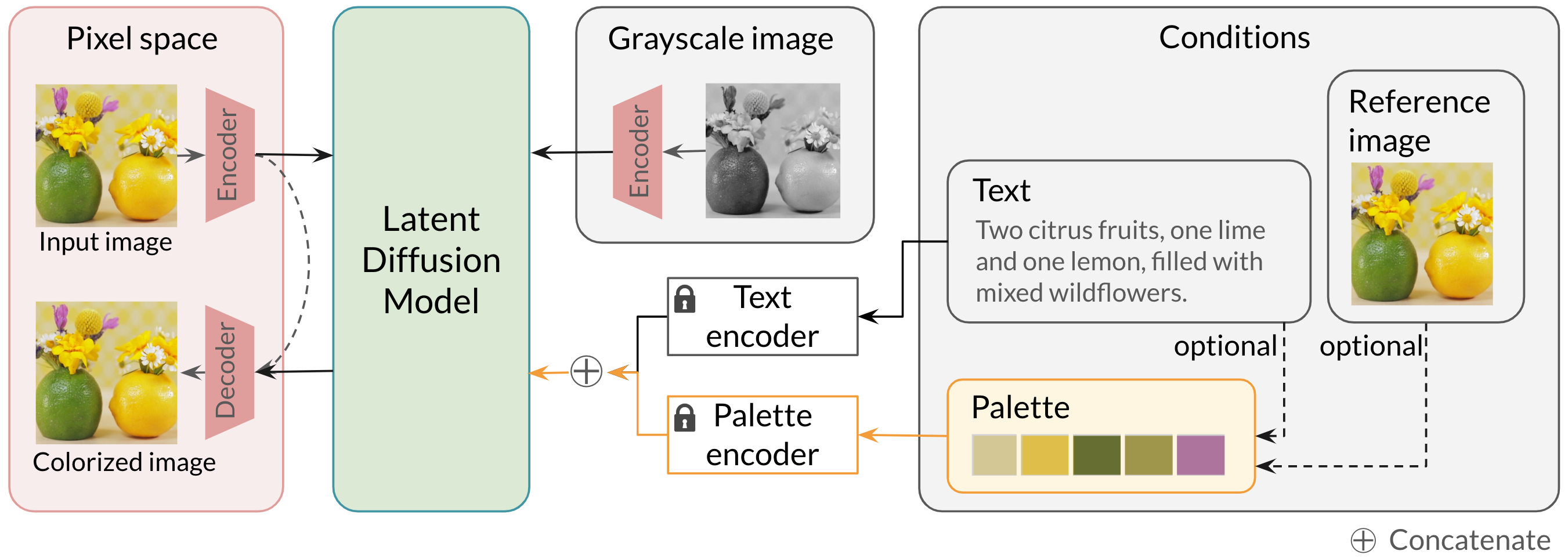}
    \caption{Palette-guided diffusion model for image colorization.}
    \label{fig:model-a}
  \end{subfigure}
  \hfill
  \begin{subfigure}{0.35\linewidth}
    \includegraphics[width=\linewidth]{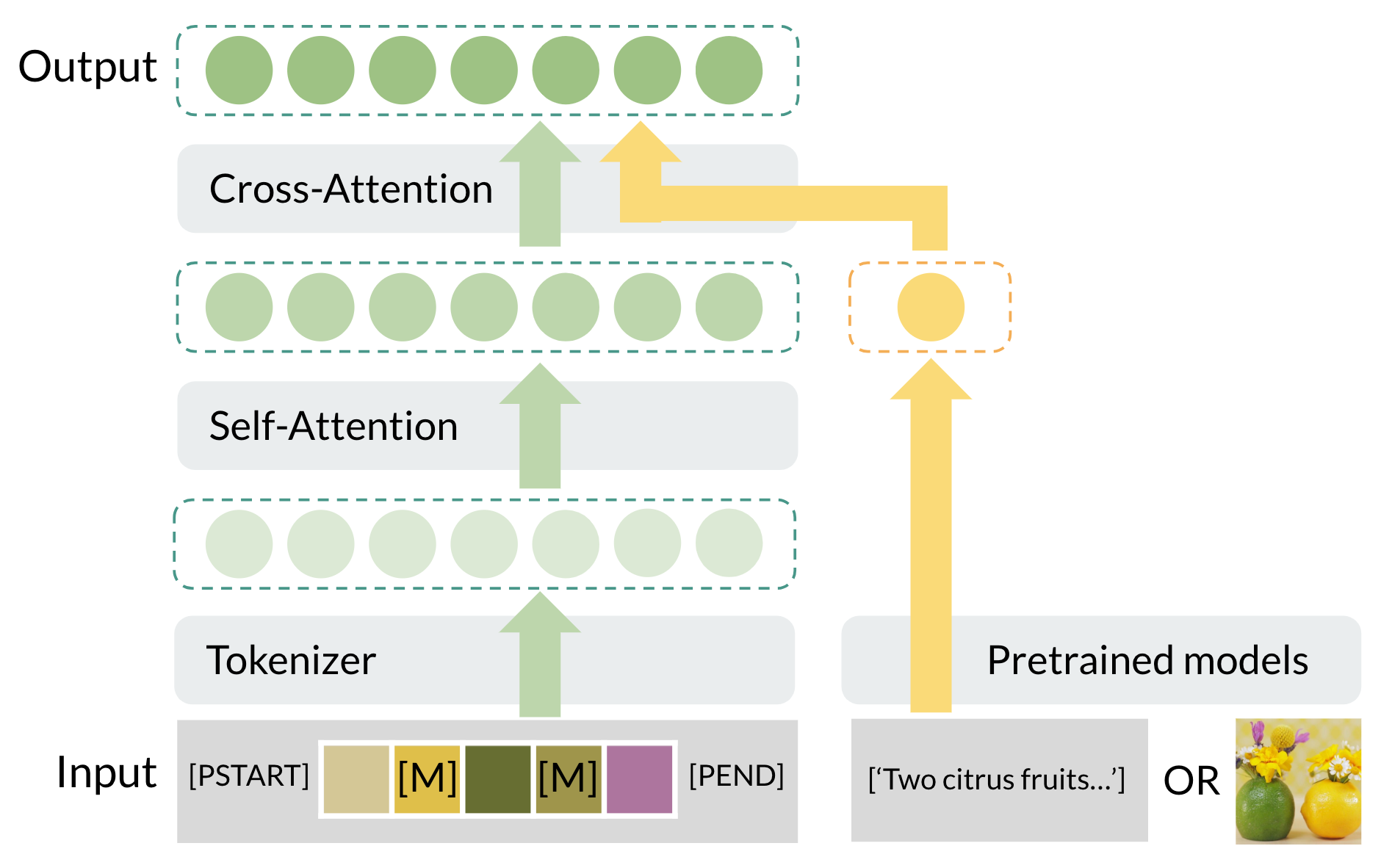}
    \caption{Masked color model for palette representation.}
    \label{fig:model-b}
  \end{subfigure}
  \vspace{-5pt}
  \caption{The framework for exploring various palette representation guidance in a diffusion-based image colorization model. The grayscale image is processed to provide structural guidance, while text and palette embeddings are fused to guide color and content. Text and reference image are optional conditions to achieve palette embeddings. During the training process, the reference image is the original input image, while a different image can be used in the inference process.}
  \label{fig:framework}
\end{figure*}

\subsection{Exemplar-based color control}
Using reference images to guide color in image generation is a widely adopted strategy, typically implemented via style transfer methods. Style transfer approaches aim to apply both the color and texture of a reference image to a target image. An early diffusion-based style transfer method, InST~\cite{zhang2023inversion}, encodes a style image into an optimized textual embedding via fine-tuning, requiring separate training for each style. StyleDiffusion~\cite{wang2023stylediffusion} improves control by disentangling style and content using a CLIP-based style loss during fine-tuning, enabling more interpretable stylization. More recent training-free methods offer greater flexibility. StyleID~\cite{chung2024style} injects style image features into the attention layers of generation networks, using AdaIN for color alignment. Cross-Image Attention~\cite{alaluf2024cross} achieves zero-shot appearance transfer by aligning semantic features between the content and style images. However, these approaches tend to entangle color with other visual attributes, such as texture or brush strokes, making it difficult to achieve precise and isolated color control. In contrast, our method focuses solely on color guidance, aiming to enable global color control without introducing unintended texture-related effects.
\section{Approach}

\subsection{Overview of palette guidance framework}

We propose a palette-guided image colorization framework built upon diffusion models, as illustrated in Figure \ref{fig:model-a}. The framework aims to generate a colorized image from a grayscale input, guided by a textual description and a color palette. These three modalities, grayscale image, text prompt, and palette, are treated as separate conditioning inputs to the diffusion model. The grayscale image provides spatial structure, while the text and palette inputs supply semantic and chromatic cues. The grayscale image is incorporated into the diffusion process via a ControlNet-inspired architecture, which serves exclusively to convey spatial structural information. Meanwhile, the palette input is fused with the textual prompt and introduced through a cross-attention-based conditioning mechanism.
This multi-modal conditioning design enables users to specify desired colors through manually selected palettes or palettes derived from reference images, alongside textual descriptions.

\subsection{Conditioning design in diffusion models}

To incorporate palette and textual guidance into diffusion-based generation process, we concatenate the palette and text embeddings with aligned dimensions to form a unified conditioning vector. This integrated conditioning vector is injected into the model via cross-attention mechanism at multiple layers of the U-Net, enabling the model to learn color semantics from the palette and content semantics from the text. 
The palette embeddings are derived from a dedicated palette representation model (see Section~\ref{palette_repsentation}), while text embeddings are obtained from the CLIP~\cite{radford2021learning} text encoder. Both embeddings are projected into the same space for joint conditioning.
In addition to the text and palette inputs, we incorporate the grayscale image as a structural condition via the ControlNet branch. While inspired by ControlNet's capacity to accept various structural inputs, in our design it serves solely to convey spatial layout from the grayscale image. This setup allows us to preserve the pretrained backbone of the diffusion model, while ensuring that palette-based color guidance remains independent from the grayscale conditioning branch.

While our current framework is demonstrated for image colorization, the palette conditioning mechanism can be flexibly applied to other color-sensitive generative tasks such as text-to-image generation or reference-based style transfer.

\subsection{Palette representation models}
\label{palette_repsentation}
A key component of our approach is the palette representation model, which generates embeddings from raw color palettes. In design, color choices are influenced by multiple factors beyond the colors themselves, including textual descriptions and reference images. To account for these factors, we explore three types of color palette representation models: a palette-only model, a text-palette model conditioned on text, and an image-palette model conditioned on a reference image. These variants allow us to analyze the impact of different modalities on palette representation and tailor our approach for downstream tasks.

\textbf{Palette-only model.}
Color selection is often guided by intuition and color theory principles, such as complementary or analogous color relationships. The palette-only model is designed to learn the intrinsic relationships among colors in a palette, focusing on the inherent harmony and balance of colors. In this model, the input is solely a color palette without any additional context.

\textbf{Text-palette model.}
Text can evoke specific color associations in people's minds. For example, the word "Christmas" might immediately bring to mind the combination of red and green. The text-palette model learns not only the relationships among colors in the palette, but also the connections between those colors and a given textual description. This model produces palette features that are both harmonious and contextually relevant to the provided text, making it particularly valuable in scenarios where color choices need to align with a theme described in text. 

\textbf{Image-palette model.}
Extracting a representative palette from an image is a common practice in design. The image-palette model is designed to learn the connections between palette colors and their source images. This model is particularly useful in scenarios where a reference image is provided, as the learned palette features help maintain a consistent color style in the generated image.

To develop these palette representation models, we extend the multimodal masked color modeling approach proposed by Qiu et al.~\cite{qiu2023multimodal}. In this approach, a color is represented as a color code and a palette as a sequence of color codes. Colors are assigned to discrete bins in the CIELAB color space using a $b \times b \times b$ histogram, with $b=16$. We prepend and append each palette sequence with special tokens \verb|[PSTART]| and \verb|[PEND]| to denote the beginning and end of a palette. If a palette is shorter than the fixed sequence length, \verb|[PAD]| tokens are added to reach the required length. During training, color codes tokens are embedded as vectors in a learned space, and a certain percentage of input tokens are randomly masked by replacing them with a \verb|[MASK]| token. 
The palette model is trained to predict the masked tokens, enabling it to learn meaningful internal representation of palettes.

Depending on the application scenario, textual and image-based conditions can be integrated with the palette representation. As shown in Figure \ref{fig:model-b}, the palette-only model encodes color tokens using a self-attention module. For the text-palette and image-palette models, color embeddings are integrated with text or image embeddings via cross-attention modules. Text and image embeddings are extracted using large pretrained models, and the final layer of network produces a palette embedding conditioned on these inputs.

In addition, we explore the performance of different large pretrained models for representing the text and image inputs. For the text-palette model, we compare text encoders from both cross-modal and text-only models. Specifically, we investigate the CLIP~\cite{radford2021learning} text encoder, which is trained on image-text pairs, and Sentence-BERT (SBERT)~\cite{Reimers2019SentenceBERTSE} which is trained solely on textural data. Similarly, for the image-palette models, we evaluate image encoders from both cross-modal and image-only models. In particular, we examine the CLIP image encoder versus DINOv2~\cite{oquab2023dinov2}, a self-supervised vision model trained only on visual data. The impact of these encoder choices is analyzed in the experimental results (Section~\ref{sec:plt_rep_analysis}).
\section{Experimental design}
\label{sec:exp}

\subsection{Datasets}

Diverse and balanced color distributions are essential for learning effective palette representations. We constructed a Palette-Text-Image dataset from LAION dataset~\cite{schuhmann2022laion5bopenlargescaledataset, schuhmann2021laion} (LAION-22k-PTI). We collected 22,783 images from LAION-400M dataset~\cite{schuhmann2021laion} using CLIP-based retrieval with the query "natural image". From each image, we extracted a five-color palette using K-means clustering in CIELAB color space. This process yielded the LAION-22k-PTI dataset with 18,226 / 2,278 / 2,279 samples for training, validation and testing. 
We also created a COCO-Stuff Palette-Text-Image dataset (COCO-Stuff-PTI) from the extended COCO-Stuff dataset~\cite{caesar2018coco}, which is widely used in image colorization research~\cite{weng2022code, chang2022coder, weng2024cad}, including 59,371 training data and 2,468 validation data.

\begin{figure}[t]
    \centering
    \includegraphics[width=1.0\linewidth]{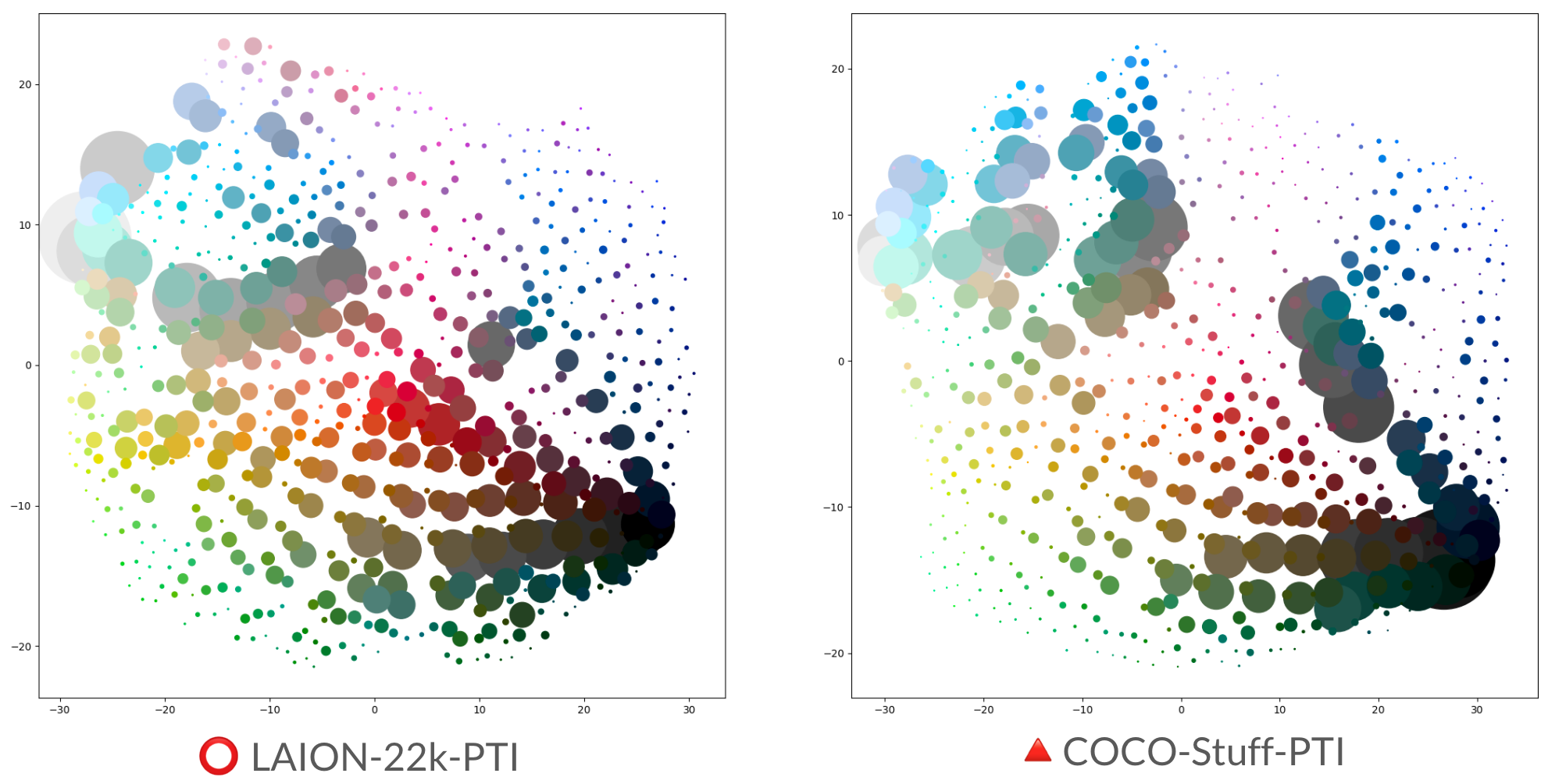}
    \caption{Color distribution of the LAION-22k-PTI and COCO-Stuff-PTI datasets, that each point is assigned its corresponding color and the size of each point reflects its frequency.}
    \label{fig:color_distibution}
\end{figure}

To visualize the color distribution of both datasets, we projected the 3D CIELAB color values of all palette colors into 2D using t-SNE (t-Distributed Stochastic Neighbor Embedding). As shown in Figure \ref{fig:color_distibution}, the LAION-22k-PTI dataset contains a boarder range of chromatic colors than the COCO-Stuff-PTI dataset.
Therefore, we used the LAION-22k-PTI datasets (18,226 training samples) to train both the palette representation model and the image colorization models. In the evaluation phase, we test on 2,279 samples from the LAION-22k-PTI test set and 2,468 samples from the COCO-Stuff-PTI validation set. The samples from the LAION-22k-PTI test set are used in user study.

\subsection{Implementation details}

\begin{table*}
  \caption{Results of palette representation models predicting 1-5 masked colors on the LAION-22k-PTI test set.}
  \label{tab:result_color_model_laion}
  {
    \begin{tabular}{ll@{\hspace{10pt}}ccccc@{\hspace{10pt}}ccccc}
    \hline
    \toprule
    Palette model & Condition & \multicolumn{5}{c}{Accuracy(\%)@1$\uparrow$} &  \multicolumn{5}{c}{Similarity: DCCW@1$\downarrow$}\\
    \cmidrule{3-7} \cmidrule{8-12}
    & Encoder & {1 color} & {2 colors} & {3 colors} & {4 colors} & {5 colors} & {1 color} & {2 colors} & {3 colors} & {4 colors} & {5 colors}\\ 
    \midrule
    palette-only & - & 12.68 & 1.78 & 0.32 & 0.22 & 0.01 & 2.69 & 5.51 & 8.45 & 11.91 & 18.52 \\ 
    \hline
    text-palette & SBERT & 14.05 & 2.45 & 0.45 & \textbf{0.17} & 0.00 & 2.53 & 5.14 & 7.99 & 11.20 & 17.00 \\
    text-palette & CLIP & \textbf{15.12} & \textbf{2.57} & \textbf{0.59} & 0.15 & 0.00 & \textbf{2.47} & \textbf{5.08} & \textbf{7.74} & \textbf{10.81} & \textbf{16.08} \\
    \hline
    image-palette & DINOv2 & 14.07 & 2.36 & 0.80 & 0.37 & \textbf{0.29} & 2.50 & 4.95 & 7.50 & 10.15 & 13.52 \\
    image-palette & CLIP & \textbf{19.96} & \textbf{4.41} & \textbf{1.22} & \textbf{0.48} & 0.07 & \textbf{2.12} & \textbf{4.22} & \textbf{6.34} & \textbf{8.53} & \textbf{11.33} \\
    \bottomrule
    \end{tabular}
  }
\end{table*}

\begin{table*}
  \caption{Results of palette representation models predicting 1-5 masked colors on the COCO-Stuff-PTI validation set.}
  \label{tab:result_color_model_coco}
  {
    \begin{tabular}{ll@{\hspace{10pt}}ccccc@{\hspace{10pt}}ccccc}
    \hline
    \toprule
    Palette model & Condition & \multicolumn{5}{c}{Accuracy(\%)@1$\uparrow$} &  \multicolumn{5}{c}{Similarity: DCCW@1$\downarrow$}\\
    \cmidrule{3-7} \cmidrule{8-12}
    & Encoder & {1 color} & {2 colors} & {3 colors} & {4 colors} & {5 colors} & {1 color} & {2 colors} & {3 colors} & {4 colors} & {5 colors}\\ 
    \midrule
    palette-only & - & 12.22 & 2.24 & 0.60 & 0.15 & 0.00 & 2.18 & 4.39 & 6.66 & 9.15 & 14.57 \\ 
    \hline
    text-palette & SBERT & 13.62 & 2.44 & 0.51 & 0.14 & 0.00 & 2.14 & 4.29 & 6.60 & 9.28 & 14.65 \\
    text-palette & CLIP & \textbf{15.29} & \textbf{2.87} & \textbf{0.64} & \textbf{0.19} & 0.00 & \textbf{2.08} & \textbf{4.24} & \textbf{6.51} & \textbf{9.11} & \textbf{14.34} \\
    \hline
    image-palette & DINOv2 & 8.88 & 1.12 & 0.11 & 0.01 & 0.00 & 2.63 & 5.33 & 8.25 & 11.35 & 15.34 \\
    image-palette & CLIP & \textbf{15.46} & \textbf{3.21} & \textbf{0.80} & \textbf{0.23} & \textbf{0.03} & \textbf{2.00} & \textbf{4.02} & \textbf{6.24} & \textbf{6.23} & \textbf{11.44} \\
    \bottomrule
    \end{tabular}
  }
\end{table*}

\begin{table*}
  \caption{Comparison of image colorization performance across methods on the LAION-22k-PTI test set.}
  \label{tab:result_colorization_laion}
  {
    \begin{tabular}{lllcccccc}\hline
    Method & Palette & Palette model & {Hist\_Bha$\downarrow$} & {DCCW$\downarrow$} & {PSNR$\uparrow$} & {SSIM$\uparrow$} & {LPIPS$\downarrow$} \\
    \hline
    ControlNet~\cite{zhang2023adding} & \ding{55} & - & 0.9756 & 20.57 & 12.41 & 0.6819 & 0.5421 \\
    L-CAD~\cite{weng2024cad} & \ding{55} & - & 0.9855 & 19.43 & 13.57 & \textbf{0.8474} & 0.3671 \\
    \hline
    Baseline1 (ControlNet + Palette in prompt) & \ding{51} & - & 0.9742 & 22.17 & 13.29 & 0.6957 & 0.5116 \\
    Baseline2 (L-CAD + Palette in prompt) & \ding{51} & - & 0.9811 & 17.68 & 12.88 & 0.8349 & 0.3585 \\
    \hline
    Ours & \ding{51} & palette-only & 0.9629 & 17.16 & 14.15 & 0.6324 & 0.4889 \\
    & \ding{51} & text-palette & 0.9574 & 16.07 & 15.22 & 0.7650 & 0.3938 \\
    & \ding{51} & image-palette & \textbf{0.9522} & \textbf{14.12} & \textbf{15.96} & 0.7737 & \textbf{0.3601} \\
    \hline
    \end{tabular}
  }
\end{table*}

\begin{table*}
  \caption{Comparison of image colorization performance across methods on the COCO-Stuff-PTI validation set.}
  \label{tab:result_colorization_coco}
  {
    \begin{tabular}{lllcccccc}\hline
    Method & Palette & Palette model & {Hist\_Bha$\downarrow$} & {DCCW$\downarrow$} & {PSNR$\uparrow$} & {SSIM$\uparrow$} & {LPIPS$\downarrow$} \\
    \hline
    ControlNet~\cite{zhang2023adding} & \ding{55} & - & 0.9543 & 16.68 & 14.38 & 0.5657 & 0.4819 \\
    L-CAD~\cite{weng2024cad} & \ding{55} & - & 0.9760 & 14.94 & 15.13 & 0.8419 & \textbf{0.3232} \\
    \hline
    Baseline1 (ControlNet + Palette in prompt) & \ding{51} & - & 0.9529 & 16.01 & 14.87 & 0.5697 & 0.4688 \\
    Baseline2 (L-CAD + Palette in prompt) & \ding{51} & - & 0.9796 & 15.93 & 15.85 & \textbf{0.8700} & 0.3303 \\
    \hline
    Ours & \ding{51} & palette-only & 0.9292 & 12.81 & 15.00 & 0.4730 & 0.4975 \\
    & \ding{51} & text-palette & 0.9373 & 12.96 & 16.59 & 0.6430 & 0.3761 \\
    & \ding{51} & image-palette & \textbf{0.9112} & \textbf{10.16} & \textbf{18.35} & 0.6684 & 0.3238 \\
    \hline
    \end{tabular}
  }
\end{table*}

\begin{figure*}[t]
    \centering
    \includegraphics[width=1.0\linewidth]{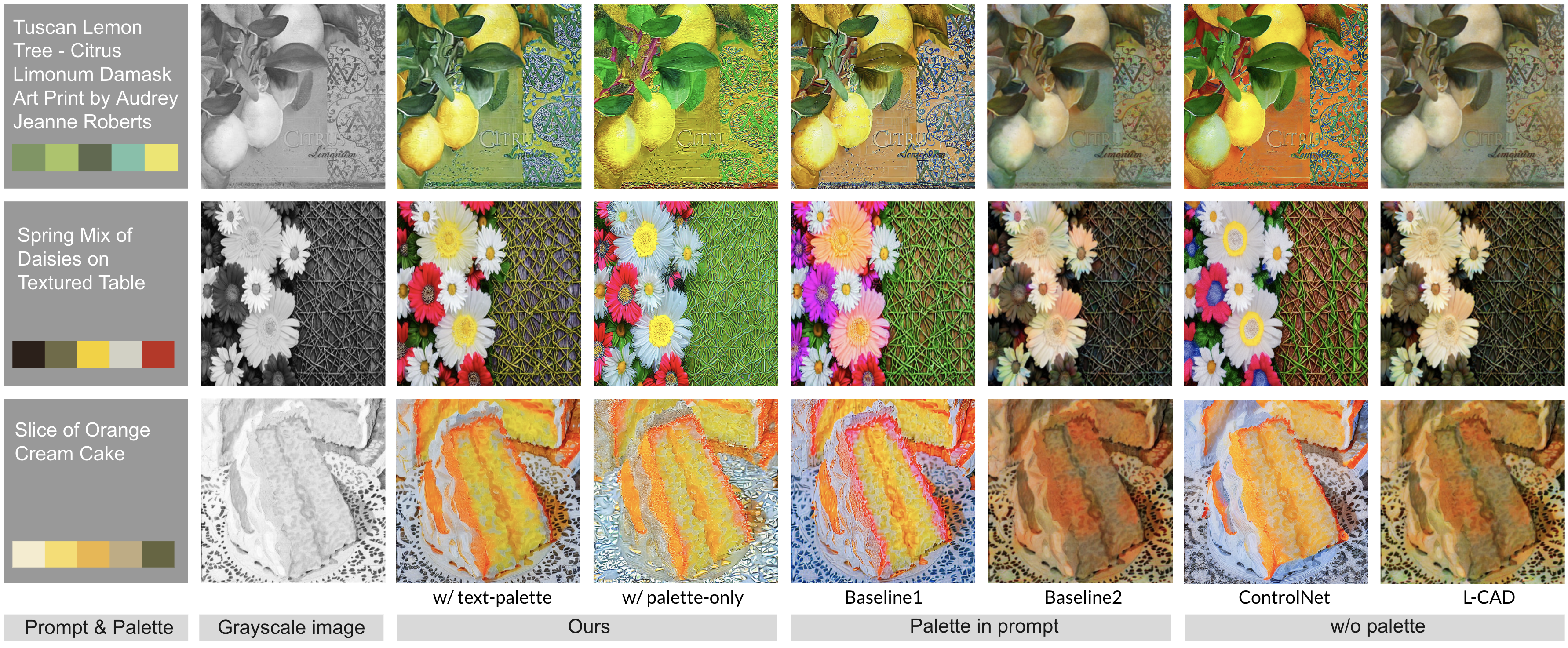}
    \caption{Comparison of colorization results using different methods. ControlNet and L-CAD refer to the original models using only text and grayscale inputs. Baseline1 and Baseline2 represent ControlNet and L-CAD with palette descriptions inserted into their text prompt. Our results using palette embeddings from palette-only and text-palette representation models.}
    \label{fig:colorization_results_t2p}
\end{figure*}

For the palette representation stage, we trained each masked color model variant on a single NVIDIA T4 GPU. In all variants of this model, each color token is transformed into a 768-dimensional vector. We employed an early stopping strategy with a patience of 30, meaning that the training process would stop if the validation loss did not improve for 30 epochs. Training each palette representation model took approximately 30 minutes. 

For the image colorization stage, we trained the ControlNet component of our framework on the Stable Diffusion 1.5 model using one NVIDIA A100 GPU. Each training epoch took about 1.5 hours with a batch size of 4. We employed with a learning rate of $1 \times 10^{-5}$ for training in the latent space. For image generation sampling, we applied the DDIM method with 50 sampling steps. Based on model performance, we selected three checkpoints of our palette-guided colorization models for final evaluation: the model with palette-only embeddings at epoch 15, the model with text-palette embeddings at epoch 20, and the model with image-palette embeddings at epoch 30. 

\subsection{Comparative methods}

\begin{figure}[t]
    \centering
    \includegraphics[width=1.0\linewidth]{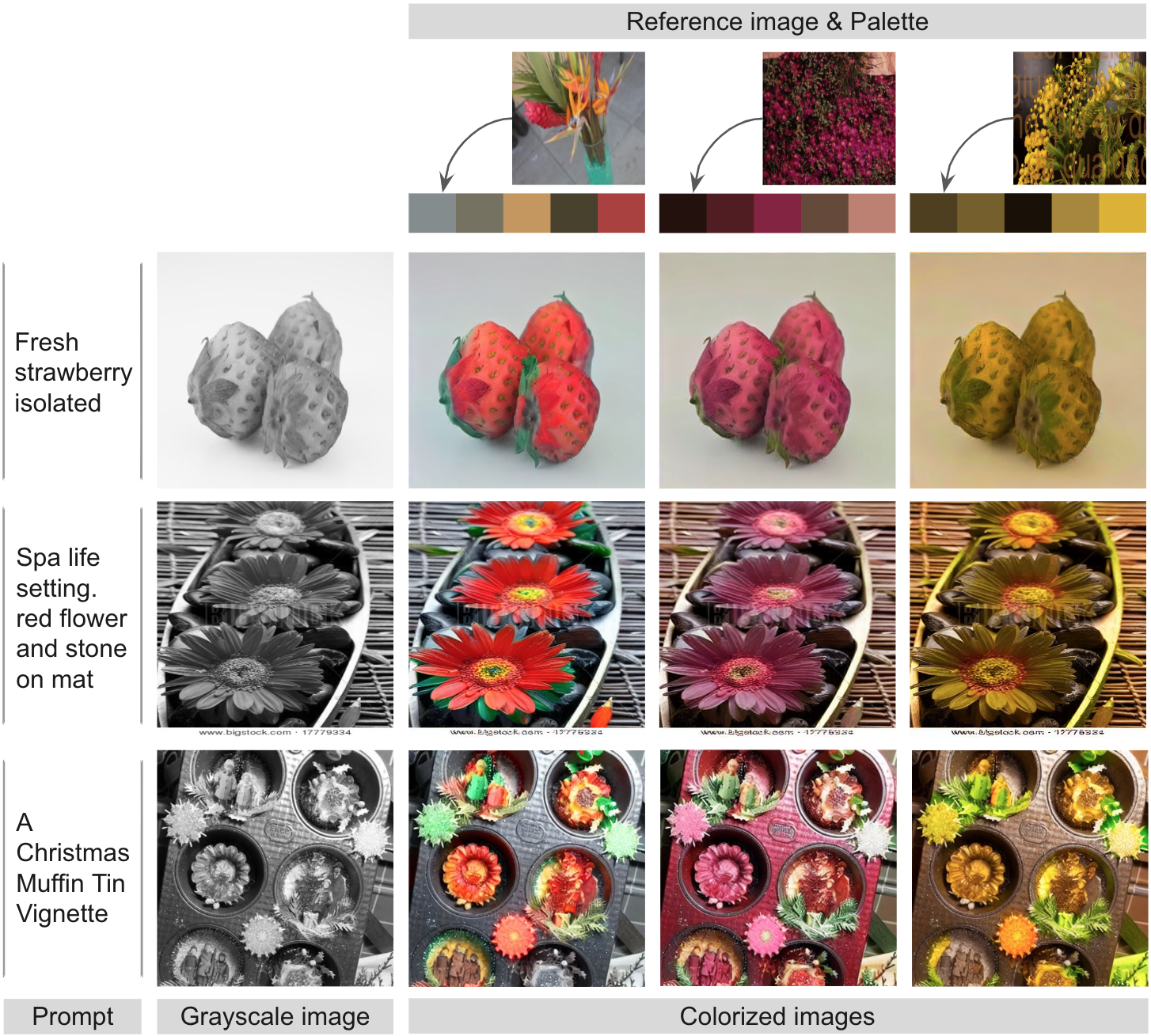}
    \caption{Colorization results of our method using palette embeddings from image-palette representation model.}
    \label{fig:colorization_results_i2p}
\end{figure}

To evaluate the effectiveness of our palette guidance approach, we selected two representative diffusion-based colorization models: ControlNet~\cite{zhang2023adding} and L-CAD~\cite{weng2024cad}. ControlNet is known for its strong spatial conditioning capabilities, while L-CAD represents the state of the art in text-driven image colorization. We first compared with both models in their original form, where only a grayscale image and text are provided, with no palette information involved.
To investigate the effect of adding palette information, we constructed two additional baselines by appending palette descriptions directly to text prompts of ControlNet and L-CAD (denoted as `+ Palette in prompt'). These prompts followed a format: "An image with a color palette of RGB(...), ..., RGB(...)". Whether using RGB or Hex notation, this simple integration yielded limited improvements.
Beyond inserting palette descriptions into text prompts, our method incorporates palette information via learned embeddings. To assess the impact of different palette representation strategies, we evaluated three versions of our model, using palette embeddings derived from a palette-only model, a text-palette model, and an image-palette model. 
To focus solely on color-related factors, none of the models employed semantic segmentation modules during testing.

\subsection{Evaluation metrics}

For evaluating the palette representation models, we assess prediction accuracy by comparing model-predicted colors against ground truth when one to five colors in a five-color palette are randomly masked. We also measure palette similarity using Dynamic Closest Color Warping (DCCW)~\cite{kim2021dynamic}, which calculates the perceptual distance between predicted and reference palettes.

For image colorization, we focus on color-level metrics that directly assess color control. We compute the Bhattacharyya distance~\cite{aherne1998bhattacharyya} to evaluate color histogram similarity (Hist\_Bha) and DCCW to compare the palette extracted from the generated image to the reference palette. These color-level metrics capture how well the generated image aligns with the intended palette.
In addition, we report common image-level metrics, including Peak Signal-to-Noise Ratio (PSNR)~\cite{huynh2008scope}, Structural Similarity Index Measure (SSIM)~\cite{wang2004image}, and Learned Perceptual Image Patch Similarity (LPIPS)~\cite{zhang2018unreasonable}, following previous colorization works~\cite{chang2022coder, chang2023coins, weng2024cad}. 
PSNR measures pixel-level differences between two images, SSIM focuses on structural similarity, and LPIPS assesses perceptual differences. 
We consider these image-level metrics as supplementary to our core goal of controllable color generation.
\section{Results and discussion}


\subsection{Palette representation analysis}
\label{sec:plt_rep_analysis}
The evaluation results for the palette representation models on the LAION-22k-PTI test set and the COCO-Stuff-PTI validation set are shown in Tables~\ref{tab:result_color_model_laion} and \ref{tab:result_color_model_coco}. Across both datasets, the text-palette and image-palette models consistently outperform the palette-only model in terms of top-1 color prediction accuracy and palette similarity, for all cases where one to five colors are masked. This demonstrates that incorporating textual and visual context significantly enhances the quality of palette representations, as the additional information enables the model to better capture and predict color relationships. Furthermore, when encoding textual conditions, cross-modal embeddings derived from CLIP outperform text-only embeddings from SBERT, indicating that CLIP's multi-modal training provides a richer understanding of the relationships between text and colors. Similarly, for encoding visual conditions, cross-modal embeddings from CLIP show superior performance compared to image-only embeddings from DINOv2, further confirming the advantage of using cross-modal features for palette guidance.

\subsection{Palette-guided Image colorization analysis}

Overall, our palette-guided models consistently outperform other methods on color-level metrics, demonstrating the effectiveness of incorporating palette information. Notably, simply inserting palette descriptions into prompts, as the baseline approaches, can even degrade performance, underscoring the importance of how palette information is integrated. Our learned palette embeddings provide more meaningful guidance, leading to more accurate and controllable colorization results. Tables~\ref{tab:result_colorization_laion} and \ref{tab:result_colorization_coco} show quantitative results on both datasets. Our palette-guided models achieve superior performance on color-level metrics, Hist\_Bha and DCCW, compared to both non-palette and palette-in-prompt baselines. These results confirm that learned palette embeddings enable more precise and consistent color control. Our method also outperforms the others on most image-level metrics such as PSNR and LPIPS. Although our SSIM scores are slightly lower than L-CAD's, this is likely due to L-CAD's explicit preservation of grayscale luminance for structural consistency. Importantly, L-CAD does not perform as well in color alignment, which remains our primary focus.

Figure~\ref{fig:colorization_results_t2p} presents qualitative comparisons, where our method takes both text and palette as input, while the baseline methods with the same inputs and others without any palette input. When prompts lack explicit color specifications for certain objects, as "Tuscan lemon tree" and "Daisies" in the first two rows, the results without palette or with palette in prompt tend to exhibit disorganized or uncontrolled color schemes. In contrast, our palette-guided methods produce results more closely to the target palette. When the text prompt includes partial color cues, such as "Orange Cream Cake" in the last row, our palette-guided model using the text-palette representation yields more precise color expressions, demonstrating a clear advantage in color control. 

Figure~\ref{fig:colorization_results_i2p} illustrates palette-guided colorization using reference images. Our method using image-palette representation effectively transfers the reference image’s color style via the extracted palette, enabling diverse outputs from the same text prompt. For example, by varying the input palette, our model can generate different realistic appearances for "Strawberries", demonstrating the flexibility and fine-grained color control enabled by our approach.

\subsection{User study}
\label{sec:user_study}
\begin{figure}[t]
    \centering
    \includegraphics[width=0.95\linewidth]{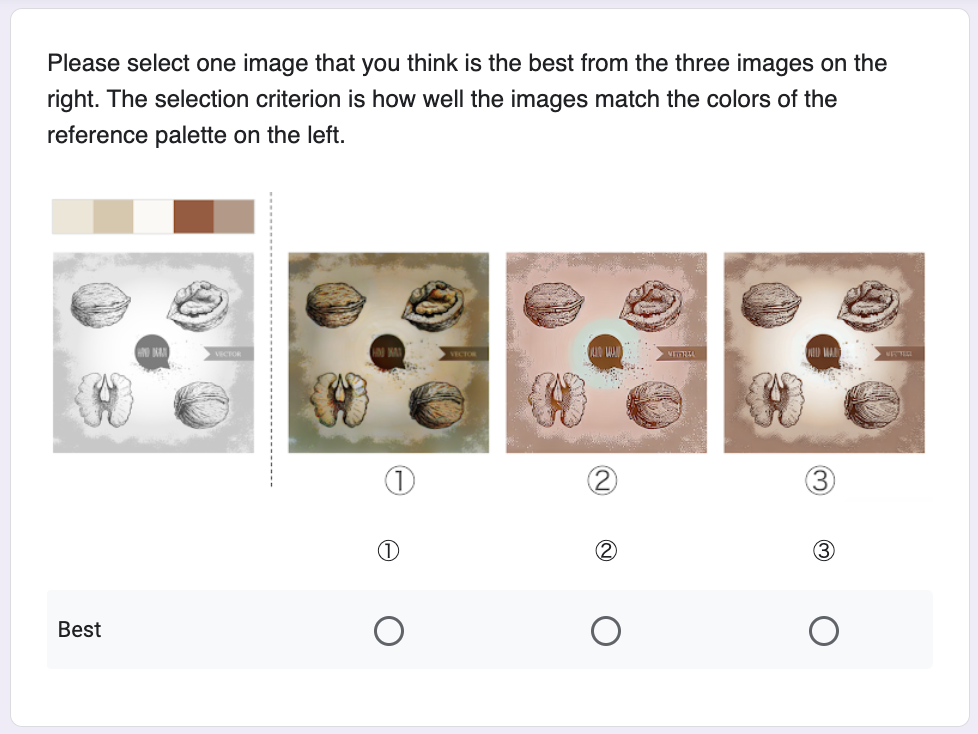}
    \caption{Evaluation interface with reference palette. Options \textcircled{\scriptsize 1}, \textcircled{\scriptsize 2} and \textcircled{\scriptsize 3} correspond to L-CAD + Palette in prompt, ControlNet + Palette in prompt, and ours using text-palette representation. The order of results is randomized.}
    \label{fig:form-1}
\end{figure}

\begin{figure}[t]
    \centering
    \includegraphics[width=0.95\linewidth]{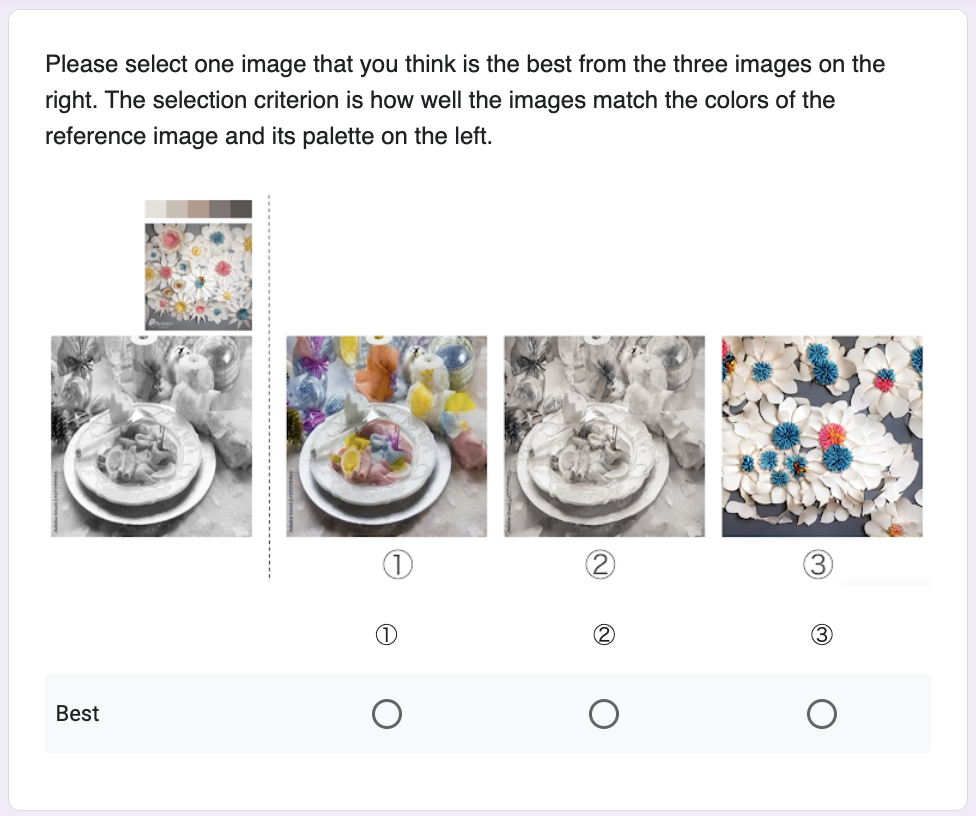}
    \caption{Evaluation interface with reference image. Options \textcircled{\scriptsize 1}, \textcircled{\scriptsize 2} and \textcircled{\scriptsize 3} correspond to ours using image-palette representation, StyleID, and Cross-Image Attention. The order of results is randomized.}
    \label{fig:form-2}
\end{figure}
\begin{table*}
  \caption{Responses from 28 designers in the interview study}
  \label{tab:eval_results}
  \begin{tabular}{ll}
    \midrule
    Q1 How useful is a color palette in design tasks?&\raisebox{-0.2\height}{\includegraphics[scale=0.3]{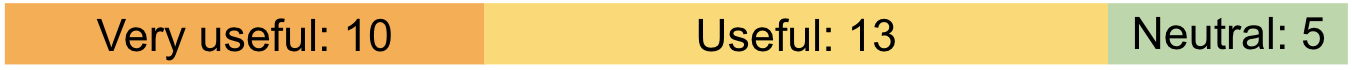}}\\
    Q2 Would you like to use color palette guidance in an image generation tool?&\raisebox{-0.2\height}{\includegraphics[scale=0.3]{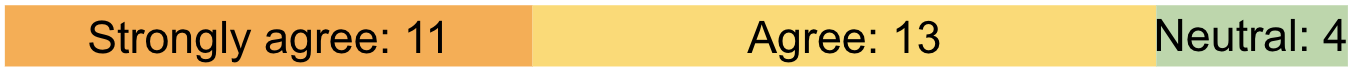}}\\
  \bottomrule
\end{tabular}
\end{table*}

We conducted a user study with professional graphic designers to further evaluate human perception for the colorized outputs of our palette-guided models. Although we developed three types of palette representation, we focused the user study on models using text-palette and image-palette representations. This focus is justified by the importance of text guidance in real-world scenarios and by our prior finding that the text-palette model consistently outperforms the palette-only variant.

Designers engaged in two evaluation scenarios, one with a reference palette and the other with a reference image, corresponding to our models guided by text-palette and image-palette representations, respectively. 
In the reference-palette scenario, we compared our model using text-palette embeddings with palette-in-prompt baselines implemented using ControlNet and L-CAD.
In the reference-image scenario, we did not include ControlNet or L-CAD, as those frameworks do not support reference image condition. Instead, we compared our model using image-palette embeddings against two recent style transfer approaches, StyleID~\cite{chung2024style} and Cross-Image Attention~\cite{alaluf2024cross}. These methods leverage reference images to guide overall visual appearance, including color and texture. The evaluation interfaces are shown in Figure~\ref{fig:form-1} and \ref{fig:form-2}.

We randomly selected 90 grayscale images from the LAION-22k-PTI test set, with 45 images allocated to each scenario. 
In both scenarios, designers were shown a grayscale input image, the corresponding reference (palette or image), and three anonymized colorized outputs (ours and other comparative methods) in random order. They were asked to select the output that best aligned with the reference in terms of color consistency.

\begin{figure}[t]
    \centering
    \includegraphics[width=1.0\linewidth]{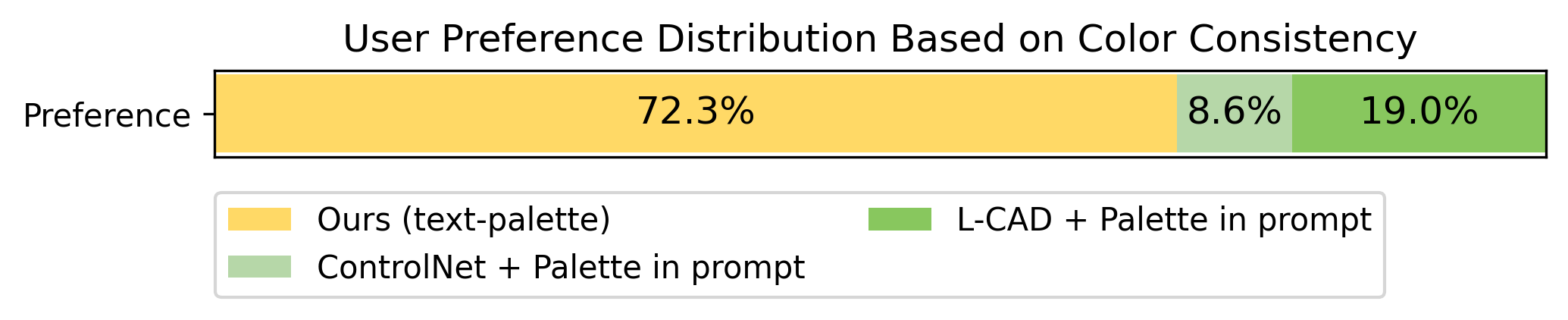}
    \caption{Evaluation results with reference palette.}
    \label{fig:eval_results_1}
\end{figure}

\begin{figure}[t]
    \centering
    \includegraphics[width=1.0\linewidth]{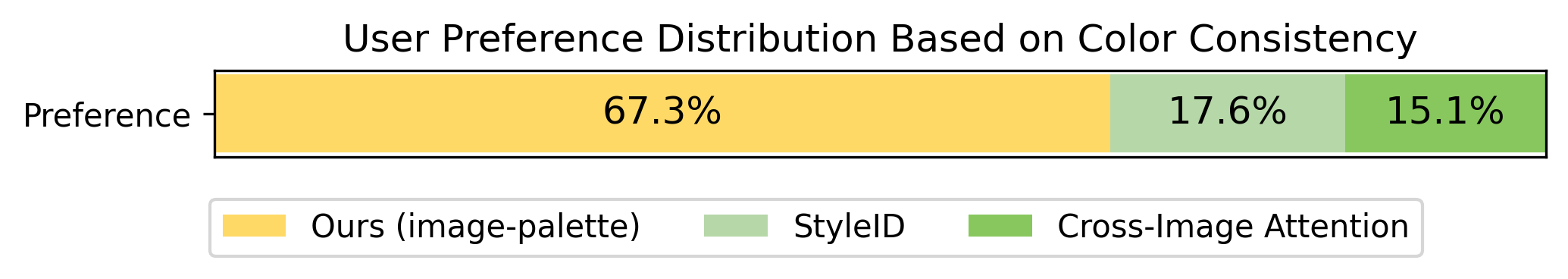}
    \caption{Evaluation results with reference image.}
    \label{fig:eval_results_2}
\end{figure}

We recruited 33 designers to participate in the study. As shown in Figure~\ref{fig:eval_results_1} and \ref{fig:eval_results_2}, our method received the highest preference in both evaluation scenarios, outperforming the palette-in-prompt baselines and recent style transfer methods in terms of color consistency. These results demonstrate the effectiveness of our palette-guided models in generating images that better match human preferences for color composition.
After the evaluation, designers completed a short questionnaire. The results shown in Table~\ref{tab:eval_results} are based on responses from 28 designers with prior experience using AI-powered generation tools. The feedback revealed a strong demand for controllable palette guidance in image generation workflows, further highlighting the practical value of our approach for real-world design applications.

\subsection{Limitations and future work}

\begin{figure}[t]
    \centering
    \includegraphics[width=1.0\linewidth]{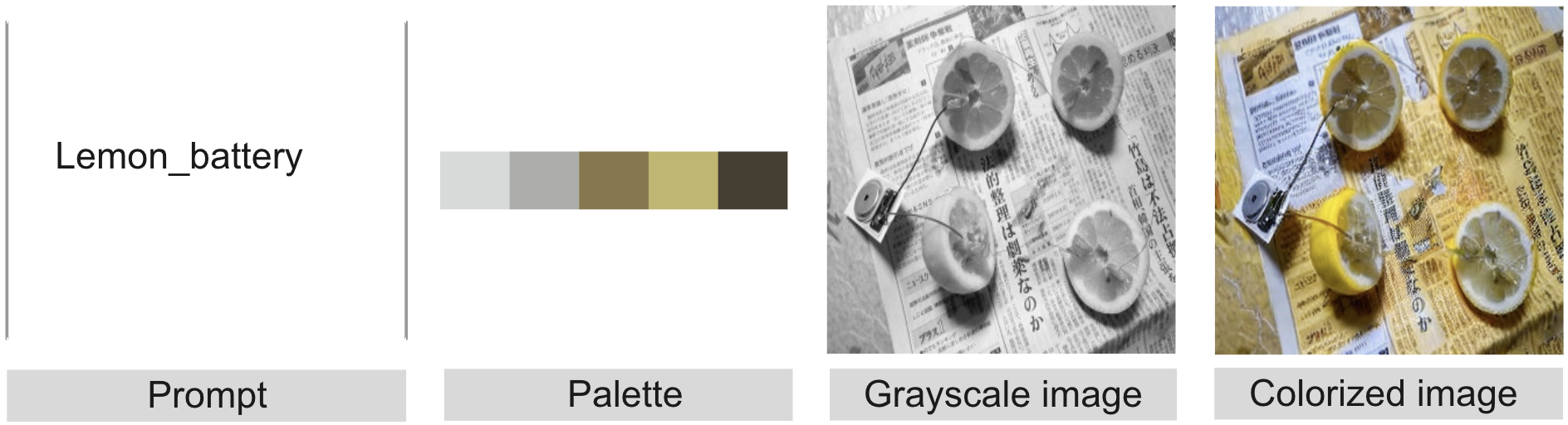}
    \caption{Failure case of color bleeding.}
    \label{fig:color_bleeding}
\end{figure}

\textbf{Color bleeding.}
In this paper, we employed image colorization as a downstream application of palette guidance, emphasizing consistency between the generated image's colors and the input palette. However, we did not specially address the issue of color bleeding, where colors may unintentionally spill over into unintended regions, as shown in Figure~\ref{fig:color_bleeding}. Solutions to this problem have been explored in other colorization studies. For example, L-CAD~\cite{weng2024cad} employs a segmentation model (e.g., SAM~\cite{kirillov2023segment}) to detect object contours and restrict colors to the appropriate regions. Incorporating such pretrained segmentation models into our framework could help reduce color bleeding by improving color boundary adherence. We consider this a promising direction for future work.

\begin{figure}[t]
    \centering
    \includegraphics[width=1.0\linewidth]{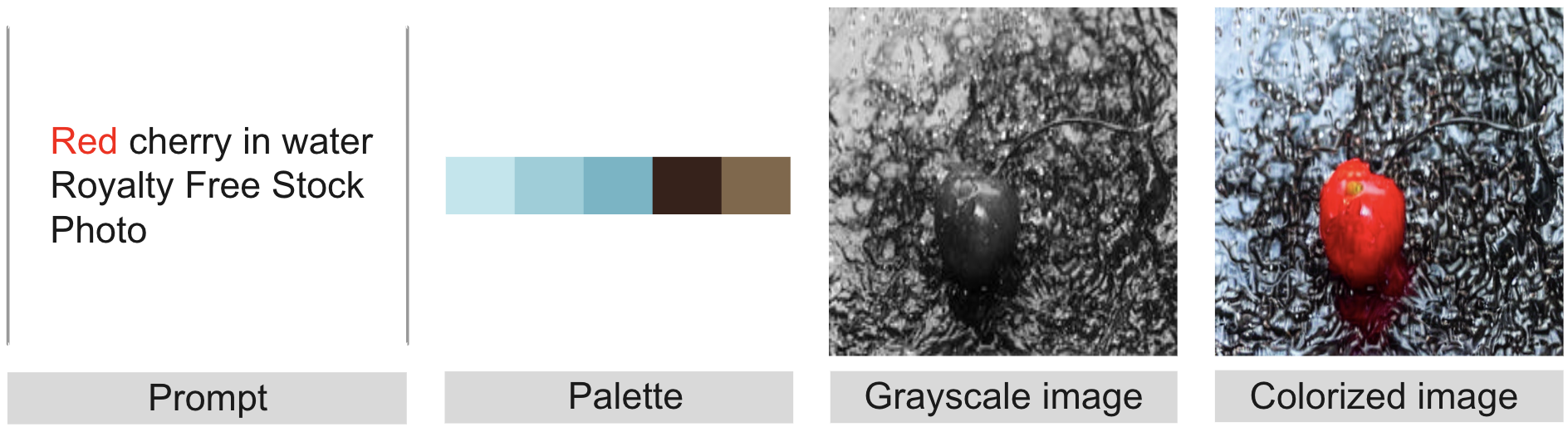}
    \caption{Conflict case between text and palette guidance.}
    \label{fig:conflict_t_p}
\end{figure}

\textbf{Conflict between text and palette guidance.}
In our training data, the color descriptions in the text prompt were consistent with the palette guidance. However, conflicts between text and palette cues can occasionally arise in practice. For example, in Figure~\ref{fig:conflict_t_p}, the colorized image reflects the red cherry specified in the text prompt, while the background colors follow the palette guidance. In this case, the palette failed to influence the cherry's color. A potential way to mitigate such conflicts is to allow users to specify a clear priority between text-based and palette-based color constraints. Additionally, developing adaptive strategies that dynamically balance text and palette guidance based on user preferences could further improve the system's handing of conflicting instructions.


\section{Conclusion}

In this paper, we introduced palette guidance alongside text prompts in diffusion-based models to enhance user control over color schemes of generated images. We developed and evaluated three types of palette representation models, palette-only, text-palette, and image-palette, and demonstrated that including textural or visual context significantly improves the accuracy and consistency of color predictions. Our experimental results show that palette-guided colorization models outperform those guided solely by text prompts, producing images that align more closely with desired color schemes, whether using user-customized palettes or palettes derived from reference images. Overall, our findings highlight the potential of palette guidance for achieving more controlled and visually coherent colorization. 
Future research could further explore the application of our palette guidance mechanism across other domains and color-related generative tasks to further enhance its adaptability and impact.

\bibliographystyle{ACM-Reference-Format}
\bibliography{base}


\begin{thebibliography}{40}


\ifx \showCODEN    \undefined \def \showCODEN     #1{\unskip}     \fi
\ifx \showISBNx    \undefined \def \showISBNx     #1{\unskip}     \fi
\ifx \showISBNxiii \undefined \def \showISBNxiii  #1{\unskip}     \fi
\ifx \showISSN     \undefined \def \showISSN      #1{\unskip}     \fi
\ifx \showLCCN     \undefined \def \showLCCN      #1{\unskip}     \fi
\ifx \shownote     \undefined \def \shownote      #1{#1}          \fi
\ifx \showarticletitle \undefined \def \showarticletitle #1{#1}   \fi
\ifx \showURL      \undefined \def \showURL       {\relax}        \fi
\providecommand\bibfield[2]{#2}
\providecommand\bibinfo[2]{#2}
\providecommand\natexlab[1]{#1}
\providecommand\showeprint[2][]{arXiv:#2}

\bibitem[Aherne et~al\mbox{.}(1998)]%
        {aherne1998bhattacharyya}
\bibfield{author}{\bibinfo{person}{Frank~J Aherne}, \bibinfo{person}{Neil~A Thacker}, {and} \bibinfo{person}{Peter~I Rockett}.} \bibinfo{year}{1998}\natexlab{}.
\newblock \showarticletitle{The Bhattacharyya metric as an absolute similarity measure for frequency coded data}.
\newblock \bibinfo{journal}{\emph{Kybernetika}} \bibinfo{volume}{34}, \bibinfo{number}{4} (\bibinfo{year}{1998}), \bibinfo{pages}{363--368}.
\newblock


\bibitem[Alaluf et~al\mbox{.}(2024)]%
        {alaluf2024cross}
\bibfield{author}{\bibinfo{person}{Yuval Alaluf}, \bibinfo{person}{Daniel Garibi}, \bibinfo{person}{Or Patashnik}, \bibinfo{person}{Hadar Averbuch-Elor}, {and} \bibinfo{person}{Daniel Cohen-Or}.} \bibinfo{year}{2024}\natexlab{}.
\newblock \showarticletitle{Cross-image attention for zero-shot appearance transfer}. In \bibinfo{booktitle}{\emph{ACM SIGGRAPH 2024 Conference Papers}}. \bibinfo{pages}{1--12}.
\newblock


\bibitem[Atif~Butt et~al\mbox{.}(2024)]%
        {atif2024colorpeel}
\bibfield{author}{\bibinfo{person}{Muhammad Atif~Butt}, \bibinfo{person}{Kai Wang}, \bibinfo{person}{Javier Vazquez-Corral}, {and} \bibinfo{person}{Joost van~de Weijer}.} \bibinfo{year}{2024}\natexlab{}.
\newblock \showarticletitle{ColorPeel: Color Prompt Learning with Diffusion Models via Color and Shape Disentanglement}.
\newblock \bibinfo{journal}{\emph{arXiv e-prints}} (\bibinfo{year}{2024}), \bibinfo{pages}{arXiv--2407}.
\newblock


\bibitem[Bojanowski et~al\mbox{.}(2017)]%
        {bojanowski2017enriching}
\bibfield{author}{\bibinfo{person}{Piotr Bojanowski}, \bibinfo{person}{Edouard Grave}, \bibinfo{person}{Armand Joulin}, {and} \bibinfo{person}{Tomas Mikolov}.} \bibinfo{year}{2017}\natexlab{}.
\newblock \showarticletitle{Enriching word vectors with subword information}.
\newblock \bibinfo{journal}{\emph{Transactions of the association for computational linguistics}}  \bibinfo{volume}{5} (\bibinfo{year}{2017}), \bibinfo{pages}{135--146}.
\newblock


\bibitem[Caesar et~al\mbox{.}(2018)]%
        {caesar2018coco}
\bibfield{author}{\bibinfo{person}{Holger Caesar}, \bibinfo{person}{Jasper Uijlings}, {and} \bibinfo{person}{Vittorio Ferrari}.} \bibinfo{year}{2018}\natexlab{}.
\newblock \showarticletitle{Coco-stuff: Thing and stuff classes in context}. In \bibinfo{booktitle}{\emph{Proceedings of the IEEE conference on computer vision and pattern recognition}}. \bibinfo{pages}{1209--1218}.
\newblock


\bibitem[Chang et~al\mbox{.}(2015)]%
        {chang2015palette}
\bibfield{author}{\bibinfo{person}{Huiwen Chang}, \bibinfo{person}{Ohad Fried}, \bibinfo{person}{Yiming Liu}, \bibinfo{person}{Stephen DiVerdi}, {and} \bibinfo{person}{Adam Finkelstein}.} \bibinfo{year}{2015}\natexlab{}.
\newblock \showarticletitle{Palette-based photo recoloring.}
\newblock \bibinfo{journal}{\emph{ACM Trans. Graph.}} \bibinfo{volume}{34}, \bibinfo{number}{4} (\bibinfo{year}{2015}), \bibinfo{pages}{139--1}.
\newblock


\bibitem[Chang et~al\mbox{.}(2022)]%
        {chang2022coder}
\bibfield{author}{\bibinfo{person}{Zheng Chang}, \bibinfo{person}{Shuchen Weng}, \bibinfo{person}{Yu Li}, \bibinfo{person}{Si Li}, {and} \bibinfo{person}{Boxin Shi}.} \bibinfo{year}{2022}\natexlab{}.
\newblock \showarticletitle{L-CoDer: Language-based colorization with color-object decoupling transformer}. In \bibinfo{booktitle}{\emph{European Conference on Computer Vision}}. Springer, \bibinfo{pages}{360--375}.
\newblock


\bibitem[Chang et~al\mbox{.}(2023)]%
        {chang2023coins}
\bibfield{author}{\bibinfo{person}{Zheng Chang}, \bibinfo{person}{Shuchen Weng}, \bibinfo{person}{Peixuan Zhang}, \bibinfo{person}{Yu Li}, \bibinfo{person}{Si Li}, {and} \bibinfo{person}{Boxin Shi}.} \bibinfo{year}{2023}\natexlab{}.
\newblock \showarticletitle{L-CoIns: Language-based colorization with instance awareness}. In \bibinfo{booktitle}{\emph{Proceedings of the IEEE/CVF Conference on Computer Vision and Pattern Recognition}}. \bibinfo{pages}{19221--19230}.
\newblock


\bibitem[Cho et~al\mbox{.}(2017)]%
        {cho2017palettenet}
\bibfield{author}{\bibinfo{person}{Junho Cho}, \bibinfo{person}{Sangdoo Yun}, \bibinfo{person}{Kyoung Mu~Lee}, {and} \bibinfo{person}{Jin Young~Choi}.} \bibinfo{year}{2017}\natexlab{}.
\newblock \showarticletitle{Palettenet: Image recolorization with given color palette}. In \bibinfo{booktitle}{\emph{Proceedings of the ieee conference on computer vision and pattern recognition workshops}}. \bibinfo{pages}{62--70}.
\newblock


\bibitem[Chung et~al\mbox{.}(2024)]%
        {chung2024style}
\bibfield{author}{\bibinfo{person}{Jiwoo Chung}, \bibinfo{person}{Sangeek Hyun}, {and} \bibinfo{person}{Jae-Pil Heo}.} \bibinfo{year}{2024}\natexlab{}.
\newblock \showarticletitle{Style injection in diffusion: A training-free approach for adapting large-scale diffusion models for style transfer}. In \bibinfo{booktitle}{\emph{Proceedings of the IEEE/CVF conference on computer vision and pattern recognition}}. \bibinfo{pages}{8795--8805}.
\newblock


\bibitem[Du et~al\mbox{.}(2021)]%
        {du2021video}
\bibfield{author}{\bibinfo{person}{Zhengjun Du}, \bibinfo{person}{Kai-Xiang Lei}, \bibinfo{person}{Kun Xu}, \bibinfo{person}{Jianchao Tan}, {and} \bibinfo{person}{Yotam~I Gingold}.} \bibinfo{year}{2021}\natexlab{}.
\newblock \showarticletitle{Video recoloring via spatial-temporal geometric palettes.}
\newblock \bibinfo{journal}{\emph{ACM Trans. Graph.}} \bibinfo{volume}{40}, \bibinfo{number}{4} (\bibinfo{year}{2021}), \bibinfo{pages}{150--1}.
\newblock


\bibitem[Gong et~al\mbox{.}(2023)]%
        {gong2023recolornerf}
\bibfield{author}{\bibinfo{person}{Bingchen Gong}, \bibinfo{person}{Yuehao Wang}, \bibinfo{person}{Xiaoguang Han}, {and} \bibinfo{person}{Qi Dou}.} \bibinfo{year}{2023}\natexlab{}.
\newblock \showarticletitle{RecolorNeRF: Layer decomposed radiance fields for efficient color editing of 3D scenes}. In \bibinfo{booktitle}{\emph{Proceedings of the 31st ACM International Conference on Multimedia}}. \bibinfo{pages}{8004--8015}.
\newblock


\bibitem[Huynh-Thu and Ghanbari(2008)]%
        {huynh2008scope}
\bibfield{author}{\bibinfo{person}{Quan Huynh-Thu} {and} \bibinfo{person}{Mohammed Ghanbari}.} \bibinfo{year}{2008}\natexlab{}.
\newblock \showarticletitle{Scope of validity of PSNR in image/video quality assessment}.
\newblock \bibinfo{journal}{\emph{Electronics letters}} \bibinfo{volume}{44}, \bibinfo{number}{13} (\bibinfo{year}{2008}), \bibinfo{pages}{800--801}.
\newblock


\bibitem[Kim et~al\mbox{.}(2022)]%
        {kim2022colorbo}
\bibfield{author}{\bibinfo{person}{Eunseo Kim}, \bibinfo{person}{Jeongmin Hong}, \bibinfo{person}{Hyuna Lee}, {and} \bibinfo{person}{Minsam Ko}.} \bibinfo{year}{2022}\natexlab{}.
\newblock \showarticletitle{Colorbo: Envisioned mandala coloringthrough human-ai collaboration}. In \bibinfo{booktitle}{\emph{Proceedings of the 27th International Conference on Intelligent User Interfaces}}. \bibinfo{pages}{15--26}.
\newblock


\bibitem[Kim and Choi(2021)]%
        {kim2021dynamic}
\bibfield{author}{\bibinfo{person}{Suzi Kim} {and} \bibinfo{person}{Sunghee Choi}.} \bibinfo{year}{2021}\natexlab{}.
\newblock \showarticletitle{Dynamic closest color warping to sort and compare palettes}.
\newblock \bibinfo{journal}{\emph{ACM Transactions on Graphics (TOG)}} \bibinfo{volume}{40}, \bibinfo{number}{4} (\bibinfo{year}{2021}), \bibinfo{pages}{1--15}.
\newblock


\bibitem[Kirillov et~al\mbox{.}(2023)]%
        {kirillov2023segment}
\bibfield{author}{\bibinfo{person}{Alexander Kirillov}, \bibinfo{person}{Eric Mintun}, \bibinfo{person}{Nikhila Ravi}, \bibinfo{person}{Hanzi Mao}, \bibinfo{person}{Chloe Rolland}, \bibinfo{person}{Laura Gustafson}, \bibinfo{person}{Tete Xiao}, \bibinfo{person}{Spencer Whitehead}, \bibinfo{person}{Alexander~C Berg}, \bibinfo{person}{Wan-Yen Lo}, {et~al\mbox{.}}} \bibinfo{year}{2023}\natexlab{}.
\newblock \showarticletitle{Segment anything}. In \bibinfo{booktitle}{\emph{Proceedings of the IEEE/CVF International Conference on Computer Vision}}. \bibinfo{pages}{4015--4026}.
\newblock


\bibitem[Kita and Miyata(2016)]%
        {kita2016aesthetic}
\bibfield{author}{\bibinfo{person}{Naoki Kita} {and} \bibinfo{person}{Kazunori Miyata}.} \bibinfo{year}{2016}\natexlab{}.
\newblock \showarticletitle{Aesthetic rating and color suggestion for color palettes}. In \bibinfo{booktitle}{\emph{Computer Graphics Forum}}, Vol.~\bibinfo{volume}{35}. Wiley Online Library, \bibinfo{pages}{127--136}.
\newblock


\bibitem[Kuang et~al\mbox{.}(2023)]%
        {kuang2023palettenerf}
\bibfield{author}{\bibinfo{person}{Zhengfei Kuang}, \bibinfo{person}{Fujun Luan}, \bibinfo{person}{Sai Bi}, \bibinfo{person}{Zhixin Shu}, \bibinfo{person}{Gordon Wetzstein}, {and} \bibinfo{person}{Kalyan Sunkavalli}.} \bibinfo{year}{2023}\natexlab{}.
\newblock \showarticletitle{Palettenerf: Palette-based appearance editing of neural radiance fields}. In \bibinfo{booktitle}{\emph{Proceedings of the IEEE/CVF Conference on Computer Vision and Pattern Recognition}}. \bibinfo{pages}{20691--20700}.
\newblock


\bibitem[O'Donovan et~al\mbox{.}(2011)]%
        {o2011color}
\bibfield{author}{\bibinfo{person}{Peter O'Donovan}, \bibinfo{person}{Aseem Agarwala}, {and} \bibinfo{person}{Aaron Hertzmann}.} \bibinfo{year}{2011}\natexlab{}.
\newblock \showarticletitle{Color compatibility from large datasets}.
\newblock In \bibinfo{booktitle}{\emph{ACM SIGGRAPH 2011 papers}}. \bibinfo{pages}{1--12}.
\newblock


\bibitem[Oquab et~al\mbox{.}(2023)]%
        {oquab2023dinov2}
\bibfield{author}{\bibinfo{person}{Maxime Oquab}, \bibinfo{person}{Timoth{\'e}e Darcet}, \bibinfo{person}{Th{\'e}o Moutakanni}, \bibinfo{person}{Huy Vo}, \bibinfo{person}{Marc Szafraniec}, \bibinfo{person}{Vasil Khalidov}, \bibinfo{person}{Pierre Fernandez}, \bibinfo{person}{Daniel Haziza}, \bibinfo{person}{Francisco Massa}, \bibinfo{person}{Alaaeldin El-Nouby}, {et~al\mbox{.}}} \bibinfo{year}{2023}\natexlab{}.
\newblock \showarticletitle{Dinov2: Learning robust visual features without supervision}.
\newblock \bibinfo{journal}{\emph{arXiv preprint arXiv:2304.07193}} (\bibinfo{year}{2023}).
\newblock


\bibitem[Podell et~al\mbox{.}(2023)]%
        {podell2023sdxl}
\bibfield{author}{\bibinfo{person}{Dustin Podell}, \bibinfo{person}{Zion English}, \bibinfo{person}{Kyle Lacey}, \bibinfo{person}{Andreas Blattmann}, \bibinfo{person}{Tim Dockhorn}, \bibinfo{person}{Jonas M{\"u}ller}, \bibinfo{person}{Joe Penna}, {and} \bibinfo{person}{Robin Rombach}.} \bibinfo{year}{2023}\natexlab{}.
\newblock \showarticletitle{Sdxl: Improving latent diffusion models for high-resolution image synthesis}.
\newblock \bibinfo{journal}{\emph{arXiv preprint arXiv:2307.01952}} (\bibinfo{year}{2023}).
\newblock


\bibitem[Qiu et~al\mbox{.}(2023a)]%
        {qiu2023multimodal}
\bibfield{author}{\bibinfo{person}{Qianru Qiu}, \bibinfo{person}{Xueting Wang}, {and} \bibinfo{person}{Mayu Otani}.} \bibinfo{year}{2023}\natexlab{a}.
\newblock \showarticletitle{Multimodal Color Recommendation in Vector Graphic Documents}. In \bibinfo{booktitle}{\emph{Proceedings of the 31st ACM International Conference on Multimedia}}. \bibinfo{pages}{4003--4011}.
\newblock


\bibitem[Qiu et~al\mbox{.}(2023b)]%
        {qiu2023color}
\bibfield{author}{\bibinfo{person}{Qianru Qiu}, \bibinfo{person}{Xueting Wang}, \bibinfo{person}{Mayu Otani}, {and} \bibinfo{person}{Yuki Iwazaki}.} \bibinfo{year}{2023}\natexlab{b}.
\newblock \showarticletitle{Color recommendation for vector graphic documents based on multi-palette representation}. In \bibinfo{booktitle}{\emph{Proceedings of the IEEE/CVF Winter Conference on Applications of Computer Vision}}. \bibinfo{pages}{3621--3629}.
\newblock


\bibitem[Radford et~al\mbox{.}(2021)]%
        {radford2021learning}
\bibfield{author}{\bibinfo{person}{Alec Radford}, \bibinfo{person}{Jong~Wook Kim}, \bibinfo{person}{Chris Hallacy}, \bibinfo{person}{Aditya Ramesh}, \bibinfo{person}{Gabriel Goh}, \bibinfo{person}{Sandhini Agarwal}, \bibinfo{person}{Girish Sastry}, \bibinfo{person}{Amanda Askell}, \bibinfo{person}{Pamela Mishkin}, \bibinfo{person}{Jack Clark}, {et~al\mbox{.}}} \bibinfo{year}{2021}\natexlab{}.
\newblock \showarticletitle{Learning transferable visual models from natural language supervision}. In \bibinfo{booktitle}{\emph{International conference on machine learning}}. PMLR, \bibinfo{pages}{8748--8763}.
\newblock


\bibitem[Reimers and Gurevych(2019)]%
        {Reimers2019SentenceBERTSE}
\bibfield{author}{\bibinfo{person}{Nils Reimers} {and} \bibinfo{person}{Iryna Gurevych}.} \bibinfo{year}{2019}\natexlab{}.
\newblock \showarticletitle{Sentence-BERT: Sentence Embeddings using Siamese BERT-Networks}. In \bibinfo{booktitle}{\emph{Conference on Empirical Methods in Natural Language Processing}}.
\newblock
\urldef\tempurl%
\url{https://api.semanticscholar.org/CorpusID:201646309}
\showURL{%
\tempurl}


\bibitem[Rombach et~al\mbox{.}(2022)]%
        {rombach2022high}
\bibfield{author}{\bibinfo{person}{Robin Rombach}, \bibinfo{person}{Andreas Blattmann}, \bibinfo{person}{Dominik Lorenz}, \bibinfo{person}{Patrick Esser}, {and} \bibinfo{person}{Bj{\"o}rn Ommer}.} \bibinfo{year}{2022}\natexlab{}.
\newblock \showarticletitle{High-resolution image synthesis with latent diffusion models}. In \bibinfo{booktitle}{\emph{Proceedings of the IEEE/CVF conference on computer vision and pattern recognition}}. \bibinfo{pages}{10684--10695}.
\newblock


\bibitem[Schuhmann et~al\mbox{.}(2022)]%
        {schuhmann2022laion5bopenlargescaledataset}
\bibfield{author}{\bibinfo{person}{Christoph Schuhmann}, \bibinfo{person}{Romain Beaumont}, \bibinfo{person}{Richard Vencu}, \bibinfo{person}{Cade Gordon}, \bibinfo{person}{Ross Wightman}, \bibinfo{person}{Mehdi Cherti}, \bibinfo{person}{Theo Coombes}, \bibinfo{person}{Aarush Katta}, \bibinfo{person}{Clayton Mullis}, \bibinfo{person}{Mitchell Wortsman}, \bibinfo{person}{Patrick Schramowski}, \bibinfo{person}{Srivatsa Kundurthy}, \bibinfo{person}{Katherine Crowson}, \bibinfo{person}{Ludwig Schmidt}, \bibinfo{person}{Robert Kaczmarczyk}, {and} \bibinfo{person}{Jenia Jitsev}.} \bibinfo{year}{2022}\natexlab{}.
\newblock \bibinfo{title}{LAION-5B: An open large-scale dataset for training next generation image-text models}.
\newblock
\showeprint[arxiv]{2210.08402}~[cs.CV]
\urldef\tempurl%
\url{https://arxiv.org/abs/2210.08402}
\showURL{%
\tempurl}


\bibitem[Schuhmann et~al\mbox{.}(2021)]%
        {schuhmann2021laion}
\bibfield{author}{\bibinfo{person}{Christoph Schuhmann}, \bibinfo{person}{Richard Vencu}, \bibinfo{person}{Romain Beaumont}, \bibinfo{person}{Robert Kaczmarczyk}, \bibinfo{person}{Clayton Mullis}, \bibinfo{person}{Aarush Katta}, \bibinfo{person}{Theo Coombes}, \bibinfo{person}{Jenia Jitsev}, {and} \bibinfo{person}{Aran Komatsuzaki}.} \bibinfo{year}{2021}\natexlab{}.
\newblock \showarticletitle{Laion-400m: Open dataset of clip-filtered 400 million image-text pairs}.
\newblock \bibinfo{journal}{\emph{arXiv preprint arXiv:2111.02114}} (\bibinfo{year}{2021}).
\newblock


\bibitem[Tan et~al\mbox{.}(2018)]%
        {tan2018efficient}
\bibfield{author}{\bibinfo{person}{Jianchao Tan}, \bibinfo{person}{Jose Echevarria}, {and} \bibinfo{person}{Yotam Gingold}.} \bibinfo{year}{2018}\natexlab{}.
\newblock \showarticletitle{Efficient palette-based decomposition and recoloring of images via RGBXY-space geometry}.
\newblock \bibinfo{journal}{\emph{ACM Transactions on Graphics (TOG)}} \bibinfo{volume}{37}, \bibinfo{number}{6} (\bibinfo{year}{2018}), \bibinfo{pages}{1--10}.
\newblock


\bibitem[Wang et~al\mbox{.}(2004)]%
        {wang2004image}
\bibfield{author}{\bibinfo{person}{Zhou Wang}, \bibinfo{person}{Alan~C Bovik}, \bibinfo{person}{Hamid~R Sheikh}, {and} \bibinfo{person}{Eero~P Simoncelli}.} \bibinfo{year}{2004}\natexlab{}.
\newblock \showarticletitle{Image quality assessment: from error visibility to structural similarity}.
\newblock \bibinfo{journal}{\emph{IEEE transactions on image processing}} \bibinfo{volume}{13}, \bibinfo{number}{4} (\bibinfo{year}{2004}), \bibinfo{pages}{600--612}.
\newblock


\bibitem[Wang et~al\mbox{.}(2023)]%
        {wang2023stylediffusion}
\bibfield{author}{\bibinfo{person}{Zhizhong Wang}, \bibinfo{person}{Lei Zhao}, {and} \bibinfo{person}{Wei Xing}.} \bibinfo{year}{2023}\natexlab{}.
\newblock \showarticletitle{Stylediffusion: Controllable disentangled style transfer via diffusion models}. In \bibinfo{booktitle}{\emph{Proceedings of the IEEE/CVF International Conference on Computer Vision}}. \bibinfo{pages}{7677--7689}.
\newblock


\bibitem[Weng et~al\mbox{.}(2022)]%
        {weng2022code}
\bibfield{author}{\bibinfo{person}{Shuchen Weng}, \bibinfo{person}{Hao Wu}, \bibinfo{person}{Zheng Chang}, \bibinfo{person}{Jiajun Tang}, \bibinfo{person}{Si Li}, {and} \bibinfo{person}{Boxin Shi}.} \bibinfo{year}{2022}\natexlab{}.
\newblock \showarticletitle{L-CoDe: Language-based colorization using color-object decoupled conditions}. In \bibinfo{booktitle}{\emph{Proceedings of the AAAI Conference on Artificial Intelligence}}, Vol.~\bibinfo{volume}{36}. \bibinfo{pages}{2677--2684}.
\newblock


\bibitem[Weng et~al\mbox{.}(2024)]%
        {weng2024cad}
\bibfield{author}{\bibinfo{person}{Shuchen Weng}, \bibinfo{person}{Peixuan Zhang}, \bibinfo{person}{Yu Li}, \bibinfo{person}{Si Li}, \bibinfo{person}{Boxin Shi}, {et~al\mbox{.}}} \bibinfo{year}{2024}\natexlab{}.
\newblock \showarticletitle{L-CAD: Language-based Colorization with Any-level Descriptions using Diffusion Priors}.
\newblock \bibinfo{journal}{\emph{Advances in Neural Information Processing Systems}}  \bibinfo{volume}{36} (\bibinfo{year}{2024}).
\newblock


\bibitem[Xiao et~al\mbox{.}(2019)]%
        {xiao2019interactive}
\bibfield{author}{\bibinfo{person}{Yi Xiao}, \bibinfo{person}{Peiyao Zhou}, \bibinfo{person}{Yan Zheng}, {and} \bibinfo{person}{Chi-Sing Leung}.} \bibinfo{year}{2019}\natexlab{}.
\newblock \showarticletitle{Interactive deep colorization using simultaneous global and local inputs}. In \bibinfo{booktitle}{\emph{ICASSP 2019-2019 IEEE International Conference on Acoustics, Speech and Signal Processing (ICASSP)}}. IEEE, \bibinfo{pages}{1887--1891}.
\newblock


\bibitem[Xu et~al\mbox{.}(2020)]%
        {xu2020stylization}
\bibfield{author}{\bibinfo{person}{Zhongyou Xu}, \bibinfo{person}{Tingting Wang}, \bibinfo{person}{Faming Fang}, \bibinfo{person}{Yun Sheng}, {and} \bibinfo{person}{Guixu Zhang}.} \bibinfo{year}{2020}\natexlab{}.
\newblock \showarticletitle{Stylization-based architecture for fast deep exemplar colorization}. In \bibinfo{booktitle}{\emph{Proceedings of the IEEE/CVF Conference on Computer Vision and Pattern Recognition}}. \bibinfo{pages}{9363--9372}.
\newblock


\bibitem[Yin et~al\mbox{.}(2021)]%
        {yin2021yes}
\bibfield{author}{\bibinfo{person}{Wang Yin}, \bibinfo{person}{Peng Lu}, \bibinfo{person}{Zhaoran Zhao}, {and} \bibinfo{person}{Xujun Peng}.} \bibinfo{year}{2021}\natexlab{}.
\newblock \showarticletitle{Yes," Attention Is All You Need", for Exemplar based Colorization}. In \bibinfo{booktitle}{\emph{Proceedings of the 29th ACM international conference on multimedia}}. \bibinfo{pages}{2243--2251}.
\newblock


\bibitem[Zabari et~al\mbox{.}(2023)]%
        {zabari2023diffusing}
\bibfield{author}{\bibinfo{person}{Nir Zabari}, \bibinfo{person}{Aharon Azulay}, \bibinfo{person}{Alexey Gorkor}, \bibinfo{person}{Tavi Halperin}, {and} \bibinfo{person}{Ohad Fried}.} \bibinfo{year}{2023}\natexlab{}.
\newblock \showarticletitle{Diffusing colors: Image colorization with text guided diffusion}. In \bibinfo{booktitle}{\emph{SIGGRAPH Asia 2023 Conference Papers}}. \bibinfo{pages}{1--11}.
\newblock


\bibitem[Zhang et~al\mbox{.}(2023b)]%
        {zhang2023adding}
\bibfield{author}{\bibinfo{person}{Lvmin Zhang}, \bibinfo{person}{Anyi Rao}, {and} \bibinfo{person}{Maneesh Agrawala}.} \bibinfo{year}{2023}\natexlab{b}.
\newblock \showarticletitle{Adding conditional control to text-to-image diffusion models}. In \bibinfo{booktitle}{\emph{Proceedings of the IEEE/CVF International Conference on Computer Vision}}. \bibinfo{pages}{3836--3847}.
\newblock


\bibitem[Zhang et~al\mbox{.}(2018)]%
        {zhang2018unreasonable}
\bibfield{author}{\bibinfo{person}{Richard Zhang}, \bibinfo{person}{Phillip Isola}, \bibinfo{person}{Alexei~A Efros}, \bibinfo{person}{Eli Shechtman}, {and} \bibinfo{person}{Oliver Wang}.} \bibinfo{year}{2018}\natexlab{}.
\newblock \showarticletitle{The unreasonable effectiveness of deep features as a perceptual metric}. In \bibinfo{booktitle}{\emph{Proceedings of the IEEE conference on computer vision and pattern recognition}}. \bibinfo{pages}{586--595}.
\newblock


\bibitem[Zhang et~al\mbox{.}(2023a)]%
        {zhang2023inversion}
\bibfield{author}{\bibinfo{person}{Yuxin Zhang}, \bibinfo{person}{Nisha Huang}, \bibinfo{person}{Fan Tang}, \bibinfo{person}{Haibin Huang}, \bibinfo{person}{Chongyang Ma}, \bibinfo{person}{Weiming Dong}, {and} \bibinfo{person}{Changsheng Xu}.} \bibinfo{year}{2023}\natexlab{a}.
\newblock \showarticletitle{Inversion-based style transfer with diffusion models}. In \bibinfo{booktitle}{\emph{Proceedings of the IEEE/CVF conference on computer vision and pattern recognition}}. \bibinfo{pages}{10146--10156}.
\newblock


\end{thebibliography}

\end{document}